\documentclass[10pt,aps,prd,superscriptaddress,nofootinbib,nobibnotes,longbibliography,floatfix,twocolumn]{revtex4-2}

\usepackage{bm}
\usepackage{mathtools,
amsmath,
amssymb,
amsfonts,
mathrsfs,
chngcntr,
multirow}

\let\cc\corresponds
\let\corresponds\relax
\usepackage{mathabx}
\let\corresponds\cc

\usepackage[utf8]{inputenc}
\usepackage[T1]{fontenc}

\usepackage{soul}

\usepackage[dvipsnames]{xcolor}
\usepackage[unicode]{hyperref}
\hypersetup{colorlinks=true, citecolor=MidnightBlue,
            linkcolor=MidnightBlue, urlcolor=MidnightBlue, linktocpage=true}
\usepackage[normalem]{ulem}

\usepackage{graphicx}

\newcommand{\D}{\mathrm{d}}
\newcommand{\leff}{L_{\rm eff}}

\begin{document}
\title{Regular BTZ black holes from an infinite tower of corrections}

\author{Pedro G. S. Fernandes}
\email{fernandes@thphys.uni-heidelberg.de}
\affiliation{Institut f\"ur Theoretische Physik, Universit\"at Heidelberg, Philosophenweg 12, 69120 Heidelberg, Germany}

\begin{abstract}
    We explore $2+1$-dimensional scalar-tensor theories derived from well-defined dimensional regularizations of the Lovelock invariants. In the limit where an infinite series of corrections is included, we obtain theories that admit fully regular black hole solutions. We analyze the properties of these regular black holes, investigate geodesics in these spacetimes, and examine the tidal forces, finding they remain finite everywhere.
\end{abstract}

\maketitle

\section{Introduction}
Despite its remarkable success in predicting and describing a wide range of astrophysical phenomena, General Relativity (GR) inevitably predicts singularities \cite{Penrose:1964wq,Hawking:1967ju,Hawking:1970zqf}, such as those at the center of black holes or in the early universe. These singularities lead to geodesically incomplete spacetimes and the divergence of physical observables, marking a fundamental limitation of the theory. While it remains uncertain whether resolving the singularity can be achieved classically, or requires a full quantum theory of gravity, one overarching conclusion is clear: GR must be modified at sufficiently high energy scales.
Indeed, from an effective field theory perspective, the Einstein-Hilbert action describing GR is merely the leading-order approximation in an infinite derivative expansion, where higher-order terms become increasingly important at high energies.

In the absence of a complete theory or mechanism capable of resolving the singularity problem, regular black holes (see Refs. \cite{Lan:2023cvz,Carballo-Rubio:2025fnc,Torres:2022twv} for reviews), which by definition lack physical singularities, have been extensively studied as potential alternatives. These models enable the exploration of black hole interiors under the assumption that spacetime geometry alone provides an adequate description. However, such solutions are typically constructed in an \textit{ad hoc} fashion: rather than emerging as solutions to fundamental field equations, they are defined geometrically, with their effective stress-energy tensor subsequently derived through the Einstein field equations \cite{1968qtr..conf...87B,Dymnikova:1992ux,Hayward:2005gi,Bambi:2013ufa}. This approach often leads to violations of standard energy conditions, while offering little insight into what underlying field theory might produce such solutions. While many of these regular black holes can formally be understood as solutions of GR coupled to non-linear electrodynamics, they require tuning the coupling constants of the theory in terms of the integration constants of the solutions \cite{Ayon-Beato:1998hmi,Bronnikov:2000vy,Ayon-Beato:2000mjt,Dymnikova:2004zc,Bronnikov:2017sgg}. Moreover, these solutions are generically unstable \cite{DeFelice:2024seu}. Therefore, regular black holes from non-linear electrodynamics are physically non-viable. Alternative approaches to try and construct regular black holes as solutions to well-defined field theories include those in \cite{Cano:2020qhy, Cano:2020ezi, Babichev:2020qpr, Baake:2021jzv, Colleaux:2017ibe, buenoRegularBlackHoles2021, buenoRegularChargedBlack2025}.

Recently, Ref. \cite{Bueno:2024dgm} proposed an infinite tower of curvature corrections as a mechanism to resolve the singularity problem, within the framework of quasi-topological gravity in higher-dimensions \cite{aguilar-gutierrezAspectsHighercurvatureGravities2023, ahmedQuintessentialQuarticQuasitopological2017, buenoFourdimensionalBlackHoles2017, buenoGeneralizedQuasitopologicalGravities2019, buenoGeneralizedQuasitopologicalGravities2022, buenoUniversalBlackHole2017, hennigarBlackHolesEinsteinian2016, hennigarGeneralizedQuasitopologicalGravity2017, morenoClassificationGeneralizedQuasitopological2023a,Moreno:2023arp}. In this framework, in the limit where an infinite series of corrections to GR is considered, regular black hole solutions naturally exist, and they can emerge dynamically as the endpoint of gravitational collapse \cite{Bueno:2024zsx,Bueno:2024eig}. An infinite tower of curvature corrections represents a particularly well-motivated approach to resolving singularities, as it naturally aligns with the effective field theory framework, and emerge in leading candidates for quantum theories of gravity, including string theory \cite{Gross:1986mw,Gross:1986iv,Grisaru:1986vi} and asymptotically safe quantum gravity \cite{Bonanno:2020bil,Knorr:2022dsx,Knorr:2018kog,Dupuis:2020fhh}. The drawback is that the framework of Ref. \cite{Bueno:2024dgm} is valid only in higher-dimensions, and the considered theories have equations of motion of higher-order for generic background, making them susceptible to Ostrograsky instability.

More recently, Ref. \cite{FernandesSingularity} explored infinite towers of regularized curvature corrections in four dimensions, leading to cosmological models where the initial singularity is replaced by an inflationary phase, as well as regular planar black holes. The theories examined in Ref. \cite{FernandesSingularity} are scalar-tensor theories within the Horndeski class that arise from a dimensional regularization procedure applied to the Lovelock invariants, a method that has gained significant attention in recent years \cite{Mann:1992ar, Glavan:2019inb, Fernandes:2020nbq, Hennigar:2020lsl, Lu:2020iav, Kobayashi:2020wqy, Fernandes:2021dsb} as a means of obtaining a Gauss-Bonnet corrected theory of gravity in four dimensions, see Ref. \cite{Fernandes:2022zrq} for a review. The novelty of Ref. \cite{FernandesSingularity} was to consider not only the Gauss-Bonnet term -- the quadratic Lovelock invariant -- but the whole tower of higher curvature corrections and the resulting phenomenology.

In this paper, we build upon the work done in Refs. \cite{hennigarLowerdimensionalGaussBonnetGravity2020, hennigarRotatingGaussBonnetBTZ2021} (see also Refs. \cite{alkac3DLovelockGravity2023,alkacMicroscopicEntropyStatic2025,konoplyaBTZBlackHoles2020,maVacuaExactSolutions2020,Skvortsova:2023zca, Alkac:2023mvr,Jusufi:2023fpo,Ahmed:2022dpu,Cuadros-Melgar:2022lrf,Dimov:2021fbm,Narzilloev:2021jtg,alkacThreeDimensionalModified2020,Alkac:2022fuc} for related work), which first explored the theories emerging from the dimensional regularization of the Gauss-Bonnet term \cite{Glavan:2019inb, Fernandes:2020nbq, Hennigar:2020lsl, Lu:2020iav, Kobayashi:2020wqy, Fernandes:2021dsb} to $2+1$ dimensions. Notably, these works found that the resulting black hole solutions closely resemble those of the higher-dimensional Einstein-Gauss-Bonnet theory \cite{Garraffo:2008hu, Boulware:1985wk, Cai:2001dz}, and represent Bañados-Teitelboim-Zanelli (BTZ) black holes \cite{Banados:1992wn,Carlip:1995qv,Banados:1992gq} corrected by the higher-derivative terms. Following Ref. \cite{FernandesSingularity}, we extend the dimensional regularization procedure to all Lovelock orders in $2+1$ spacetime dimensions. In the limit where an infinite tower of regularized Lovelock corrections is included, we obtain everywhere regular BTZ black holes. Remarkably, these black holes feature an extremal inner horizon, which prevents the classical mass inflation instabilities that typically afflict most regular black hole models \cite{Carballo-Rubio:2022kad, Carballo-Rubio:2024dca, Franzin:2022wai}. Thus, beyond ensuring a fully regular geometry, without singularities of any kind, the higher-derivative corrections also naturally evade mass inflation instabilities. This stands in stark contrast to the cases of i) the static BTZ black hole of GR, which is geodesically incomplete \cite{Cruz:1994ir}, and the rotating BTZ black hole which has a standard Cauchy horizon, being prone to mass inflation instability \cite{Bhattacharjee:2020gbo}; and ii) most other $2+1$-dimensional black holes, where modifications of GR generally introduce curvature singularities or non-extremal Cauchy horizons \cite{hennigarLowerdimensionalGaussBonnetGravity2020, hennigarRotatingGaussBonnetBTZ2021,Bueno:2022lhf, Bergshoeff:2009aq, Bergshoeff:2009hq,Gullu:2010pc,Paulos:2010ke, Sinha:2010ai, Oliva:2009ip, Ayon-Beato:2009rgu, Ayon-Beato:2014wla, Nam:2010dd, Clement:2009gq, Alkac:2016xlr, Gurses:2019wpb, Chernicoff:2024lkj,Martinez:1999qi,Hirschmann:1995he, Dias:2002ps,Chan:1994qa,Martinez:1996gn,Baake:2020tgk}.

The paper is structured as follows. In Sec. \ref{sec:theory}, we review the dimensional regularization procedure introduced in Refs. \cite{Fernandes:2020nbq, Hennigar:2020lsl, Lu:2020iav, Kobayashi:2020wqy, Fernandes:2021dsb}, extending it to all orders in the Lovelock invariants \cite{Colleaux:2020wfv} and adapting it to three dimensions. In Sec. \ref{sec:regularbhs}, we solve the field equations of the theories derived in Sec. \ref{sec:theory}, obtaining regular black hole solutions and analyzing their properties. Section \ref{sec:geodesics} is dedicated to studying geodesics in these spacetimes and the tidal forces experienced by observers. In Sec. \ref{sec:rotating}, we generalize the static solutions from Sec. \ref{sec:regularbhs} to the rotating case. Finally, we present our conclusions in Sec. \ref{sec:conclusions}.  

\section{Scalar-Tensor Theories from Regularized Lovelock Gravity}
\label{sec:theory}
Lovelock gravity \cite{Lovelock:1971yv, Padmanabhan:2013xyr} is a natural generalization of GR to higher dimensions. It is the unique local, diffeomorphism-invariant, purely metric theory that extends the Einstein-Hilbert action by including specific higher-order curvature terms known as Lovelock invariants, while ensuring second-order field equations. As a result, these theories avoid Ostrogradski instabilities. However, in a spacetime with $D$ dimensions, only up to $\lceil D/2 \rceil$ Lovelock invariants contribute non-trivially to the gravitational action, with higher-order terms being either topological or vanishing. This imposes a significant restriction in three and four-dimensional theories, where the Lovelock action reduces to the Einstein-Hilbert term with a cosmological constant.

In recent years, there has been growing interest in theories incorporating Gauss-Bonnet corrections in lower dimensions. This research gained momentum with Ref. \cite{Glavan:2019inb}, which proposed that, by applying a singular rescaling to the higher-dimensional Einstein-Gauss-Bonnet theory in a four-dimensional limit, one could obtain black hole and cosmological solutions closely resembling those of the higher-dimensional theory. This idea led to numerous follow-up studies. However, it was later demonstrated that this approach is ill-defined \cite{Gurses:2020ofy}. Subsequently, Refs. \cite{Fernandes:2020nbq, Hennigar:2020lsl, Lu:2020iav, Kobayashi:2020wqy, Fernandes:2021dsb} showed that a similar procedure -- originally introduced in Ref. \cite{Mann:1992ar} to recover GR dynamics in $D=2$ -- could be consistently implemented, resulting in four-dimensional scalar-tensor theories belonging to the Horndeski class \cite{Horndeski:1974wa, Kobayashi:2019hrl}. Below, we apply this regularization procedure to the whole Lovelock tower of invariants \cite{Colleaux:2020wfv}.

\begin{table*}
    \centering
\[
\begin{array}{|c|c|c|c|c|}
    \hline
    c_n & G_2 - 2/L^2 & G_3 & G_4 - 1 & f(r) \\
    \hline
    \hline
    1 & - \frac{8\ell^2 X^2}{(1-2\ell^2 X)^2} & 8\ell^2 X \frac{1-\ell^2 X}{(1-2\ell^2 X)^2} & \frac{\ell^2 X}{-1+2 \ell^2 X} - \frac{3}{2}\sqrt{2\ell^2 X} \tanh^{-1}\left( \sqrt{2\ell^2 X} \right) & \frac{r^2 f_{\rm BTZ}}{r^2 - \ell^2 f_{\rm BTZ}}\\
    \hline
    \frac{1}{n} & \frac{4X}{-1+2\ell^2 X} - \frac{2}{\ell^2} \log(1-2\ell^2 X) & \frac{4\ell^2 X}{1-2\ell^2 X} & -\sqrt{2\ell^2 X} \tanh^{-1}\left(\sqrt{2\ell^2 X}\right) & \frac{r^2}{\ell^2} \left( \exp\left(\frac{\ell^2}{r^2} f_{\rm BTZ}\right) -1\right) \\
    \hline
    \frac{1-(-1)^n}{2n} & \frac{4X}{-1+4\ell^4 X^2} + \frac{2}{\ell^2} \tanh^{-1}(2\ell^2 X) & \frac{8\ell^4 X^2}{1-4\ell^4 X^2} & \sqrt{\frac{\ell^2 X}{2}} \left(\tan^{-1}\left(\sqrt{2\ell^2 X}\right)-\tanh^{-1}\left(\sqrt{2\ell^2 X}\right) \right) & \frac{r^2}{\ell^2} \tanh\left(\frac{\ell^2}{r^2} f_{\rm BTZ}\right) \\
    \hline
    \frac{\Gamma(\frac{n}{2}) \delta_{0,k}}{\sqrt{\pi}\Gamma(\frac{n+1}{2})} & -\frac{16\ell^4 X^3}{(1-4\ell^4 X^2)^{3/2}} & \frac{2}{(1-4\ell^4 X^2)^{3/2}} & {}_{2}F_{1}\left( -\frac{1}{4}, \frac{3}{2}, \frac{3}{4}, 4\ell^4 X^2 \right) & \frac{r^2 f_{\rm BTZ}}{\sqrt{r^4 + \ell^4 f_{\rm BTZ}^2}} \\
    \hline
\end{array}
\]
\caption{Examples of re-summed Horndeski theories, together with their metric function for a black hole. In the last row, $k=(n-1)\bmod 2$, $\delta$ is the Kronecker delta.}
\label{tab:table}
\end{table*}

The $n^{\rm th}$ order Lovelock invariant is given by
\begin{equation}
    \mathcal{R}^{(n)} \equiv \frac{1}{2^n} \delta^{\mu_1 \nu_1 \dots \mu_n \nu_n}_{\alpha_1 \beta_1 \dots \alpha_n \beta_n} \prod_{i=1}^{n} R^{\alpha_i \beta_i}_{\phantom{\alpha_i \beta_i} \mu_i \nu_i},
\end{equation}
where $\delta^{\mu_1 \nu_1 \dots \mu_n \nu_n}_{\alpha_1 \beta_1 \dots \alpha_n \beta_n} \equiv n! \delta^{\mu_1}_{[\alpha_1} \delta^{\nu_1}_{\beta_1} \dots \delta^{\mu_n}_{\alpha_n} \delta^{\nu_n}_{\beta_n]}$ is the generalized Kronecker delta, and $n\geq 0$.
The dimensional regularization procedure of Refs. \cite{Fernandes:2020nbq, Hennigar:2020lsl} relies on the use of two conformally related metrics, $\tilde{g}_{\mu \nu} = e^{-2\phi} g_{\mu \nu}$, and applying the following limit at each Lovelock order:  
\begin{equation}
    \sqrt{-g}\mathcal{L}^{(n)} = \lim_{d\to 2n} \frac{\sqrt{-g}\mathcal{R}^{(n)} - \sqrt{-\tilde{g}}\tilde{\mathcal{R}}^{(n)}}{d-2n},
    \label{eq:limit}
\end{equation}
where the expressions inside the limit are evaluated in $d$ dimensions, and the limit is taken as $d$ approaches the critical dimension $2n$, where the $n^{\rm th}$ Lovelock invariant becomes topological. The tilde denotes quantities constructed from the conformal metric, up to total divergences. The subtraction of the conformal terms in Eq. \eqref{eq:limit} serves as a counterterm necessary to removing divergences that would otherwise arise. This approach was explicitly worked out for $n=1$ in $2D$, and $n=2$ in $4D$ in Refs. \cite{Mann:1992ar,Fernandes:2020nbq,Hennigar:2020lsl}. Moreover, it can be shown that the limit remains well-defined for any $n$, leading to a scalar-tensor theory with a scalar field $\phi$ and second-order equations of motion \cite{Colleaux:2020wfv,FernandesSingularity}. Once a well-defined scalar-tensor expression for each $\mathcal{L}^{(n)}$ is obtained, it can be incorporated into a theory of arbitrary dimensionality. This approach was first applied outside the critical Lovelock dimension for the $n=2$ case in Refs. \cite{hennigarLowerdimensionalGaussBonnetGravity2020, hennigarRotatingGaussBonnetBTZ2021}, where regularized Gauss-Bonnet theories in $2+1$ dimensions were explored. Notably, the black hole solutions they obtained closely resemble their higher-dimensional counterparts, reinforcing their interpretation as solutions to a three-dimensional Einstein-Gauss-Bonnet theory. In the following, we consider only three-dimensional theories.

The result of the limit \eqref{eq:limit} is, at each order $n$, when evaluated in three-dimensions, and up to an overall constant factor, given by a Horndeski Lagrangian
\begin{equation}
    \begin{aligned}
		\mathcal{L}^{(n)} =& G_2^{(n)}(\phi,X)-G_3^{(n)}(\phi,X)\Box\phi + G_4^{(n)}(\phi,X)R
		\\&
        +G_{4X}^{(n)}  \left[(\Box\phi)^2-\left(\nabla_\mu \nabla_\nu \phi\right)^2\right],
    \end{aligned}
	\label{eq:Horndeski}
\end{equation}
with functions given by
\begin{equation}
    \begin{aligned}
        &G_2^{(n)} = -2^{n+1}(n-1) X^n,\\&
        G_3^{(n)} = 2^{n} n X^{n-1},\\&
        G_4^{(n)} = -\frac{2^{n-1} n}{2n-3} X^{n-1},
        \label{eq:HorndeskiFunctions}
    \end{aligned}
\end{equation}
where $X = -\partial_\mu \phi \partial^\mu \phi/2$.
The gravitational action we consider in this work can be interpreted as a $3D$-Lovelock theory, and is given by
\begin{equation}  
    S = \int \mathrm{d}^3 x \sqrt{-g} \left[ \frac{2}{L^2} + R + \frac{1}{\ell^2}\sum_{n=2}^{\infty} c_n \ell^{2n} \mathcal{L}^{(n)} \right],  
    \label{eq:action}  
\end{equation}  
where we consider a negative cosmological constant $\Lambda = -1/L^2$, defined in terms of the anti de Sitter (AdS) radius $L$, $\{c_n\}$ are dimensionless coupling constants, and $\ell$ is a new length scale where beyond-GR effects become important, and that limits the spacetime curvature for the regular black hole solutions discussed in Sec. \ref{sec:regularbhs}.
When the series in Eq. \eqref{eq:action} is truncated at $n=2$, the resulting theory coincides, up to field and coupling redefinitions, with the one studied in Refs. \cite{hennigarLowerdimensionalGaussBonnetGravity2020, hennigarRotatingGaussBonnetBTZ2021}. The action \eqref{eq:action} belongs to the class of $3D$ Horndeski theories (see Refs. \cite{Lecoeur:2024kwe, Kobayashi:2011nu} for the full field equations) and possesses a shift symmetry in the scalar field, meaning it remains invariant under $ \phi \to \phi + C $ for any constant $ C $. As a result, there exists a conserved Noether current $ j^\mu $, whose divergence-free condition, $ \nabla_\mu j^\mu = 0 $, is equivalent to the scalar field equation. The explicit form of the Noether current can be determined using the results of Ref. \cite{Saravani:2019xwx}.

When the infinite series of corrections is included, the theory in Eq. \eqref{eq:action} can be resummed into a Horndeski theory \eqref{eq:Horndeski} with an inherently non-local structure. This non-locality is expected, as most ultraviolet (UV) completions of gravity, including string theory, feature a non-local gravitational sector. Tab. \ref{tab:table} provides explicit examples.

\section{Regular BTZ Black Holes and a Birkhoff Theorem}
\label{sec:regularbhs}
In this section, we obtain black hole solutions for the theory described by Eq. \eqref{eq:action}. We assume a circularly symmetric line element of the form
\begin{equation}
    \D s^2 = - f(t,r) N(t,r)^2 \D t^2 + \frac{\D r^2}{f(t,r)} + r^2 \D \varphi^2,
    \label{eq:metric}
\end{equation}
where $f(t,r)$ and $N(t,r)$ are functions to be determined by the field equations. In this coordinate system the BTZ black hole is given by $N(t,r)=1$, and
\begin{equation}
    f(t,r) \equiv f_{\rm BTZ}(r) = \frac{r^2}{L^2} - \mu,
    \label{eq:btz}
\end{equation}
where $\mu$ is an integration constant. For the solution to describe a black hole, $\mu$ must be positive, and it is associated with the black hole's mass. When $\mu > 0$, the spacetime exhibits a conical-like singularity in its causal structure at $r = 0$, resembling the behavior found in the Taub-NUT spacetime \cite{Banados:1992gq}. The static BTZ black hole is thus geodesically incomplete \cite{Cruz:1994ir}.

As in the planar $4D$ case examined in Ref. \cite{FernandesSingularity}, the scalar field profile  
\begin{equation}  
    \phi = \log(r/r_0),  
    \label{eq:sf}  
\end{equation}  
where $r_0$ is an arbitrary integration, automatically satisfies the scalar field equation, regardless of the order $n$. By employing this scalar field profile, the $tr$ field equation and a combination of the $tt$ and $rr$ field equations are solved at each order by imposing
\begin{equation}
    \partial_r N = 0, \quad \partial_t f = 0,
\end{equation}
which implies that $f(t,r)\equiv f(r)$ and $N(t,r)\equiv N(t)$. With a redefinition of the $t$ coordinate we can set $N=1$, without loss of generality. Finally, the remaining independent field equation can be integrated into an algebraic relation that determines $f(r)$
\begin{equation}
    \frac{1}{\ell^2} \sum_{n=1}^{\infty} c_n \ell^{2n} \left(- \frac{f(r)}{r^2} \right)^n = -\frac{f_{\rm BTZ}(r)}{r^2},
    \label{eq:regularbhs}
\end{equation}
where $f_{\rm BTZ}(r)$ is defined in Eq. \eqref{eq:btz}, and $c_1=1$. The BTZ solution is trivially recovered when the series is truncated at $n=1$, and the solution of Refs. \cite{hennigarLowerdimensionalGaussBonnetGravity2020, hennigarRotatingGaussBonnetBTZ2021} when truncated at $n=2$. Notoriously, Eq. \eqref{eq:regularbhs} is exactly the equation that determines planar black holes in higher-dimensional Lovelock gravity \cite{Cadoni:2016hhd,Hennigar:2017umz}. This correspondence arises because the class of theories considered in Eq. \eqref{eq:action}, obtained via the limiting procedure in Eq. \eqref{eq:limit}, can also be derived through a regularized Kaluza-Klein reduction of Lovelock gravity with a flat internal space \cite{Lu:2020iav, Kobayashi:2020wqy}. In this dimensional reduction, the scalar field $\phi$ corresponds to the logarithm of the scale factor of the internal space. Both the $2+1$-dimensional black holes discussed here and higher-dimensional planar black holes possess a planar (i.e., flat) base manifold geometry, with a scale factor given by $r$. By adopting the scalar field profile in Eq. \eqref{eq:sf} and considering a planar base manifold, we preserve these symmetries of the original higher-dimensional Lovelock theory prior to the reduction, and we expect the same functional form for the solutions.

As a result, the black holes described by the theory in Eq. \eqref{eq:action} satisfy a Birkhoff-type theorem: when the scalar field takes the form given in Eq. \eqref{eq:sf}, the metric \eqref{eq:metric} must be static and is fully characterized by the metric function $f(r)$, reducing to  
\begin{equation}  
    \D s^2 = - f(r) \D t^2 + \frac{\D r^2}{f(r)} + r^2 \D \varphi^2.  
    \label{eq:metric_static}  
\end{equation}  
However, this does not imply uniqueness. In particular, the BTZ geometry remains a solution to the field equations when the scalar field is constant. This mirrors the behavior of purely metric modified gravity theories in $2+1$ dimensions, which always admit the BTZ black hole as a solution \cite{Gurses:2019wpb}.

For spacetimes described by the metric in Eq. \eqref{eq:metric_static}, there are only two independent curvature invariants that can be constructed from the Riemann tensor, the metric, and their contractions. These can be chosen to be
\begin{equation}
    \begin{aligned}
        &R = -\frac{2f'(r) + r f''(r)}{r},\\&
        S^{\mu}_{\phantom{\mu}\nu} S^{\nu}_{\phantom{\nu}\mu} = \frac{(f'(r) - r f''(r))^2}{6r^2},
    \end{aligned}
\end{equation}
where $S_{\mu \nu} = R_{\mu \nu} -\frac{1}{3} g_{\mu \nu} R$, is the traceless Ricci tensor.
An analysis of the independent curvature invariants and the metric itself shows that, for the black hole to be regular near $r = 0$, the metric function must satisfy either $f(r) = \mathcal{O}(r^2)$ or $f(r) = 1 + \mathcal{O}(r^2)$. Additionally, from Eq. \eqref{eq:regularbhs}, it can be verified that if the series is truncated at $n = n_{\rm max}$, the metric function near $r = 0$ behaves as
\begin{equation}
    f(r) \approx - \left( \frac{\mu}{c_{n_{\rm max}}} \right)^{1/n_{\rm max}} \left( \frac{r^2}{\ell^2} \right)^{1-1/n_{\rm max}}.
\end{equation}
As a result, regular solutions of the form $f(r) = \mathcal{O}(r^2)$ are only possible when an infinite tower of corrections is included ($n_{\rm max} \to \infty$), and
\begin{equation}
    0 < \lim_{n \to \infty} (c_n)^{1/n} < \infty.
    \label{eq:coupling_condition}
\end{equation}
The condition in Eq. \eqref{eq:coupling_condition}, imposes that series in Eq. \eqref{eq:regularbhs} has a finite radius of convergence.

\begin{figure}[]
	\centering
	\includegraphics[width=\linewidth]{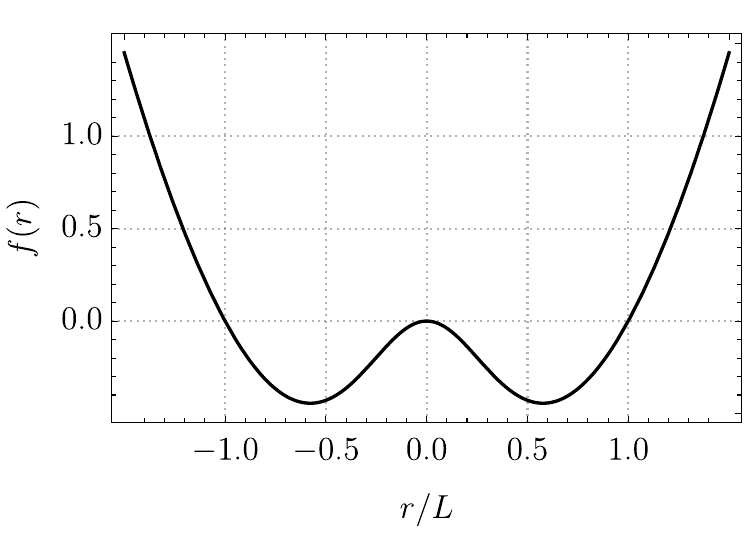}\hfill
    \caption{Regular black hole metric in Eq. \eqref{eq:metric_example} corresponding to the choice $c_n = 1$, with $\mu=1$, $L=1$ and $\ell=1/2$.}
	\label{fig:metric}
\end{figure}

There are infinitely many possible solutions to Eq. \eqref{eq:regularbhs}, depending on the couplings $\{c_n\}$ that define the theory. In Tab. \ref{tab:table}, we provide some examples of re-summed theories with couplings that satisfy the condition in Eq. \eqref{eq:coupling_condition}, along with their corresponding metric functions $f(r)$. As a representative example for the remainder of this paper, we consider the theory where $c_n=1$, which gives the following black hole solution
\begin{equation}
    f(r) = \frac{r^2 f_{\rm BTZ}}{r^2 - \ell^2 f_{\rm BTZ}},
    \label{eq:metric_example}
\end{equation}
where $f_{\rm BTZ}$ is given in Eq. \eqref{eq:btz}. The metric is regular everywhere and free of singularities of any kind, as long as $L>\ell$. Near the origin, it behaves as $f(r) = -r^2/\ell^2 + \mathcal{O}(r^4)$. In Fig. \ref{fig:metric}, we plot this metric function for a specific set of parameters, showing that the metric remains regular throughout. Other metric functions listed in Tab. \ref{tab:table} exhibit qualitatively similar shapes and regularity properties. It is important to note that all solutions are symmetric under $r \to -r$. The event horizon in all cases is located at $r_+ = L\sqrt{\mu}$, similar to the BTZ black hole. Asymptotically, the spacetime of the regular black hole behaves like AdS, with $f(r) \sim -r^2/L_{\rm eff}^2$, where the effective AdS radius is given by $L_{\rm eff}^2 = L^2 - \ell^2$ for the metric in Eq. \eqref{eq:metric_example}.

The solutions studied in this work satisfy $f(0)=0$ and $f'(0)=0$. Then, as discussed in Ref. \cite{buenoRegularChargedBlack2025}, $r=0$ corresponds to an inner horizon with vanishing surface gravity, making it extremal. This horizon is a Killing horizon associated with the Killing vector $\partial_t$. These inner-extremal horizons emerge without the need for fine-tuning and are a generic feature of the regular solutions of the theory \eqref{eq:action}. They are particularly interesting because classical mass inflation instabilities affect Cauchy horizons with non-zero surface gravity, but can be avoided if the surface gravity vanishes \cite{Carballo-Rubio:2022kad}, as occurs naturally in this case.

The solutions satisfy thermodynamic relations that exhibit the universal behavior observed in planar-horizon solutions of Lovelock gravity \cite{Hennigar:2017umz,Cadoni:2016hhd,Brenna:2015pqa}
\begin{equation}
    \D M =  T \D S, \qquad M = \frac{1}{2} T S,
\end{equation}
where
\begin{equation}
    M = \frac{\mu}{8}, \quad T = \frac{r_+}{2\pi L^2}, \quad S = \frac{\pi r_+}{2},
\end{equation}
are interpreted as the mass, Hawking temperature and entropy of the black hole, respectively. The entropy follows an area law, $S=A/4$, where $A$ is the area of the event horizon. Curiously, the same exact relations are obeyed by the BTZ black hole \eqref{eq:btz}.

We emphasize that, despite some qualitative similarities between the solutions presented here and those in Refs. \cite{buenoRegularBlackHoles2021, buenoRegularChargedBlack2025}, the underlying theories, spacetime geometries, equations of motion, and mechanism curing the singularity are fundamentally different. In particular, the theory defined by Eq. \eqref{eq:action} ensures second-order field equations, in contrast to the higher-order equations generally present in Refs. \cite{buenoRegularBlackHoles2021, buenoRegularChargedBlack2025}.

\section{Geodesics and tidal forces}
\label{sec:geodesics}
In this section, we examine the geodesics and tidal forces experienced by neutral, free-falling particles as they move through the regular black hole backgrounds discussed in the previous section. Our approach closely follows Ref. \cite{buenoRegularChargedBlack2025}. 

The metric \eqref{eq:metric_static} possesses two Killing vectors, $\partial_t$ and $\partial_\varphi$, which implies that the energy $E$ and angular momentum $J$ of particles are conserved along a geodesic. These quantities are given by
\begin{equation}
    E = f(r) \dot t, \qquad J = r^2 \dot \varphi,    
    \label{eq:conserved}
\end{equation}
where the overdot denotes a derivative with respect to the affine parameter $\lambda$, chosen such that it coincides with the proper time for a timelike geodesic. Normalization of the four-velocity of the particle $u^\mu = \dot x^\mu$, imposes
\begin{equation}
    -\epsilon = -f(r) \dot t^2 + \frac{\dot r^2}{f(r)} + r^2 \dot \varphi^2,
    \label{eq:fourvelocitynorm}
\end{equation}
where $\epsilon=0$ for a massless particle, and $\epsilon=1$ for a massive one. Using the relations of Eq. \eqref{eq:conserved} in Eq. \eqref{eq:fourvelocitynorm} we obtain
\begin{equation}
    \dot r^2 = V_{\rm eff}(r),
\end{equation}
where
\begin{equation}
    V_{\rm eff}(r) = E^2 - f(r) \left( \epsilon + \frac{J^2}{r^2} \right).
\end{equation}
is the effective potential governing the motion of the particles. Massive particles are unable to escape the AdS barrier, where $V_{\rm eff} = 0$, since the effective potential asymptotically behaves as $V_{\rm eff} \to -r^2/L_{\rm eff}^2$. In contrast, massless particles can escape the AdS barrier, provided their energies satisfy the condition $E^2 > J^2/L_{\rm eff}^2$. For the regular black holes studied in this work, the effective potential is everywhere regular and smooth.

\begin{figure}[]
	\centering
	\includegraphics[width=\linewidth]{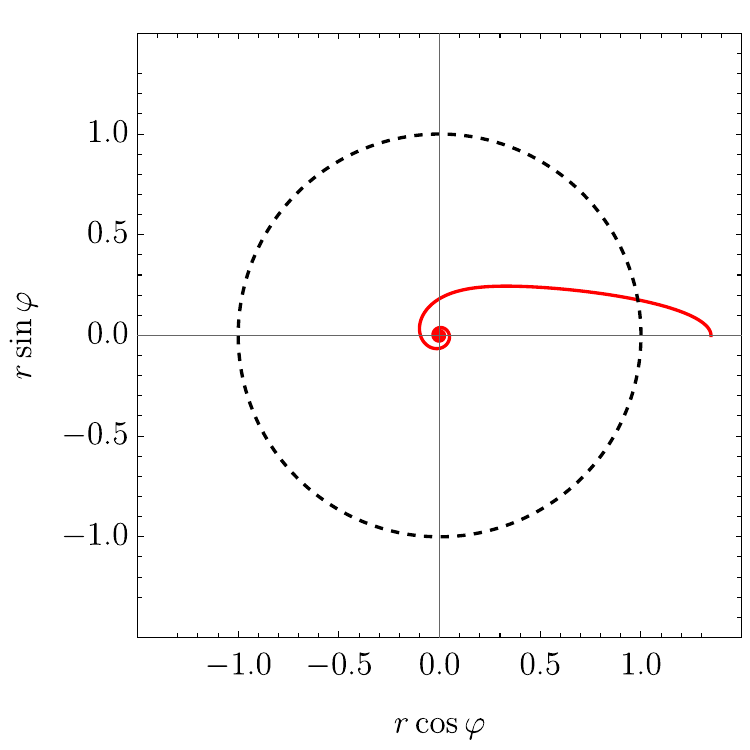}\hfill
	\includegraphics[width=\linewidth]{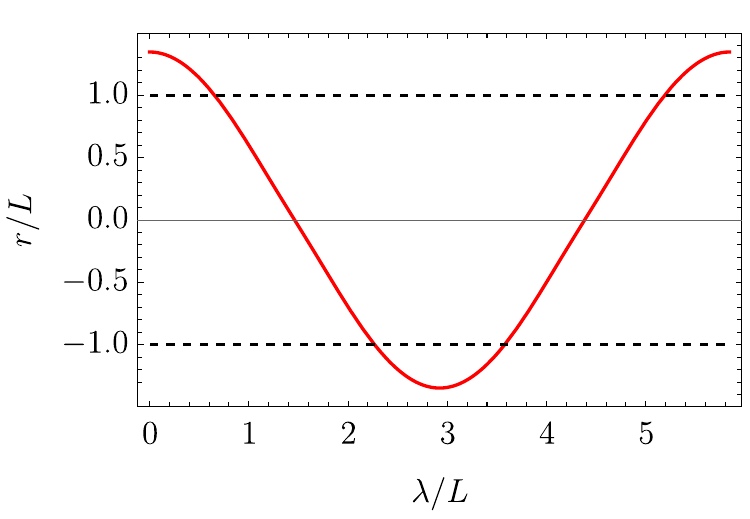}\vfill
    \caption{Timelike geodesic ($\epsilon=1$) for the metric $c_n=1$, with parameters $\mu=1$, $L=1$, $\ell=1/2$, $E^2=1$, $J=2/5$. The used initial condition is $r\approx 1.3475$. The dashed lines indicate the location of the event horizon. Top: Trajectory in the $(r \cos \varphi, r \sin \varphi)$ plane, where we observe a spiraling as $r\to 0$. Bottom: Cyclic motion of the timelike geodesic, oscilating between causally disconnected universes.}
	\label{fig:geodesic}
\end{figure}

In Fig. \ref{fig:geodesic} (top), we show an example of a timelike geodesic in the background given by Eq. \eqref{eq:metric_example} for a specific set of parameters. The particle's trajectory typically spirals around $r=0$. This behavior arises because the geodesic equations impose the relation
\begin{equation}
    \left( \frac{\D r}{\D \varphi} \right)^2 = \frac{r^2}{J^4} V_{\rm eff}.
\end{equation}
As the particle approaches $r \to 0$, and for a metric function that behaves as $f(r) \sim r^2$ in this limit (such as in Eq. \eqref{eq:metric_example}), we find that $r \sim 1/\varphi$. Therefore, as $r \to 0$, we have $\varphi \to \infty$, resulting in the spiraling behavior observed in Fig. \ref{fig:geodesic}.

Geodesics do not terminate at $r=0$, contrarily to the standard BTZ black hole \cite{Cruz:1994ir}; instead, they cross the inner-extremal horizon and move toward negative values of $r$ \cite{Zhou:2022yio}, engaging in a cyclic motion between antipodal points in causally disconnected universes. This behavior is illustrated in Fig. \ref{fig:geodesic} (bottom), where we show one complete cycle of a timelike geodesic.

To ensure that a free-falling observer is not torn apart by the tidal forces near the center of the black hole and can actually traverse to another universe, we compute the tidal forces experienced by such an observer. Denoting by $\eta^{\mu}$ a spacelike vector that represents the separation between two infinitesimally close particles following timelike geodesics, their relative acceleration is given by \cite{Lima:2020wcb, LimaJunior:2022gko}
\begin{equation}
    \frac{D^2 \eta^{\hat \alpha}}{D \lambda^2} = K^{\hat \alpha}_{\phantom{\alpha} \hat \beta} \eta^{\hat \beta},
\end{equation}
where
\begin{equation}
    K^{\hat \alpha}_{\phantom{\alpha} \hat \beta} = R^{\mu}_{\phantom{\mu} \nu \rho \sigma} e^{\hat \alpha}_{\mu} e^{\nu}_{\hat 0} e^{\rho}_{\hat 0} e^{\sigma}_{\hat \beta},
\end{equation}
is the tidal tensor, $D/D\lambda \equiv u^\mu \nabla_\mu$, is the covariant derivative along the geodesics, and $e^{\mu}_{\hat \alpha}$ form an orthonormal tetrad basis for the frame following a timelike geodesic. The non-trivial components of the tidal tensor are given by \cite{buenoRegularChargedBlack2025}
\begin{equation}
    \begin{aligned}
        &K^{\hat r}_{\phantom{\alpha} \hat r} = -\frac{1}{2}f''(r) + \frac{J^2}{2r^2} \left( -f''(r) + \frac{f'(r)}{r} \right),\\&
        K^{\hat \varphi}_{\phantom{\alpha} \hat \varphi} = -\frac{1}{2r}f'(r).
    \end{aligned}
\end{equation}
Using these expressions, along with the metric function \eqref{eq:metric_example} and those presented in Tab. \ref{tab:table}, it can be shown that the tidal forces remain non-divergent everywhere. For instance, near $r=0$, for the metric function \eqref{eq:metric_example}, we have
\begin{equation}
    \begin{aligned}
        &K^{\hat r}_{\phantom{\alpha} \hat r} = \frac{1}{\ell^2} - \frac{4J^2}{\ell^4 \mu} + \mathcal{O}\left(r^2\right),\\&
        K^{\hat \varphi}_{\phantom{\alpha} \hat \varphi} = \frac{1}{\ell^2} + \mathcal{O}\left(r^2\right),
    \end{aligned}
\end{equation}
which are finite.

\section{Rotating Black Holes}
\label{sec:rotating}
The static solutions \eqref{eq:metric_static} can be generalized to rotating ones by boosting in the $t-\varphi$ plane (see e.g. Refs. \cite{Dias:2002ps, buenoRegularBlackHoles2021, hennigarRotatingGaussBonnetBTZ2021, Hale:2024zvu})
\begin{equation}
    t\to \gamma t - \leff \omega \varphi, \quad \varphi \to \gamma \varphi - \frac{\omega}{\leff} t,
    \label{eq:boost}
\end{equation}
where $\gamma$ and $\omega$ are constant parameters related by
\begin{equation}
    \gamma^2 - \omega^2 = 1,
\end{equation}
such that the static metric is recovered when $\omega = 0$. The parameter $\omega$ is therefore directly related to the spin of the black hole. The length scale $L_{\rm eff}$ is the effective AdS radius of the spacetime, and it is employed in the boost in Eq. \eqref{eq:boost} to ensure the metric does not rotate asymptotically \cite{hennigarRotatingGaussBonnetBTZ2021}. The rotating metric is given by
\begin{equation}
    \D s^2 = - \frac{r^2}{\rho^2}f(r) \D t^2 + \frac{\D r^2}{f(r)} + \rho^2 \left(\D \varphi - N^{\varphi} \D t \right)^2,
    \label{eq:metricrot}
\end{equation}
where
\begin{equation}
    \begin{aligned}
        &\rho^2 = r^2 + \omega^2 \left( r^2 - \leff^2 f(r) \right),\\&
        N^\varphi = -\frac{\gamma \omega \left( r^2 - \leff^2 f(r) \right) }{\leff \rho^2},
    \end{aligned}
\end{equation}
and $f(r)$ is obtained from the static solutions. The scalar field remains unchanged compared to the static case and is given by Eq. \eqref{eq:sf}.

The determinant of the induced metric on constant-$r$ hypersurfaces is given by $-r^2 f(r)$. As a result, the horizon structure remains unchanged, with the event horizon located at $r_+ = L\sqrt{\mu}$ and the inner-extremal horizon at $r_- = 0$ in this coordinate system. These black holes also possess ergoregions when $g_{tt} > 0$, where the Killing vector $\partial_t$ becomes spacelike.

We note that the inner horizon of these regular solutions is extremal only due to the presence of the length scale $\ell$, as the rotating BTZ black hole is recovered in the limit $\ell \to 0$. Thus, in this case, the same mechanism that resolves the singularity also prevents mass inflation instabilities, even though these typically arise at much lower curvature scales than the singularity.

\section{Conclusions}
\label{sec:conclusions}
In this work, we have derived and discussed scalar-tensor theories resulting from the dimensional regularization of Lovelock invariants to three dimensions. This regularization procedure to three dimensions, which was first used in Refs. \cite{hennigarLowerdimensionalGaussBonnetGravity2020, hennigarRotatingGaussBonnetBTZ2021, alkac3DLovelockGravity2023,alkacMicroscopicEntropyStatic2025,konoplyaBTZBlackHoles2020,maVacuaExactSolutions2020,Skvortsova:2023zca, Alkac:2023mvr,Jusufi:2023fpo,Ahmed:2022dpu,Cuadros-Melgar:2022lrf,Dimov:2021fbm,Narzilloev:2021jtg,alkacThreeDimensionalModified2020,Alkac:2022fuc} to regularize the quadratic (Gauss-Bonnet) and cubic Lovelock terms, has been extended in this work to all orders in curvature in the Lovelock tower. The resulting theory is presented in Eq. \eqref{eq:action}.

We have demonstrated that, in the limit where an infinite tower of corrections is considered, it is possible to obtain theories with a non-local structure (see Tab. \ref{tab:table}) that admit regular black holes. These black holes obey a Birkhoff-type theorem and feature an inner horizon that is extremal. As a result, we expect these black holes to be immune to classical mass inflation instabilities, and this is achieved without any fine-tuning. Additionally, the tidal forces experienced by observers are everywhere finite, and timelike geodesics exhibit cyclic motion, oscillating between causally disconnected universes.
We have also derived rotating black hole solutions within these theories. In contrast to the BTZ black hole, these solutions do not exhibit constant curvature, a feature more akin to realistic black holes like the Kerr black hole. Furthermore, they share many physical characteristics with the Kerr black hole, such as an ergoregion, making them valuable toy models for studying more realistic scenarios.

There are several promising avenues for future research. A more in-depth exploration of the thermodynamics of these black holes, both static and rotating, would be particularly valuable. It would be interesting to investigate whether a microscopic derivation of the semiclassical black hole entropy can be achieved, following the approach of Refs. \cite{Strominger:1997eq,Correa:2010hf,Correa:2011dt}. Notably, such a derivation has already been successfully carried out for the theory containing up to cubic Lovelock corrections \cite{alkacMicroscopicEntropyStatic2025}. Additionally, we anticipate that the class of theories studied in this work satisfies a holographic c-theorem \cite{Freedman:1999gp,Sinha:2010ai,Paulos:2010ke,Paulos:2012xe,Bueno:2022lhf}, similar to the case of up-to-cubic Lovelock corrections \cite{alkac3DLovelockGravity2023}. Therefore, it would be worthwhile to determine whether the full theory described by Eq. \eqref{eq:action} also satisfies a holographic c-theorem.

The regular black holes obtained in this work feature an inner-extremal horizon with zero surface gravity. An important avenue for future research would be to investigate their dynamical stability against mass inflation; we expect no instability exists. Additionally, these black holes satisfy a Birkhoff-type theorem, implying that the exterior region of any matter configuration where the metric is of the form given in Eq. \eqref{eq:metric} will be described by the corresponding static black hole solution. This raises the interesting question of whether these regular black holes can form dynamically. Exploring gravitational collapse scenarios similar to those studied in Refs. \cite{Bueno:2024zsx, Bueno:2024eig} could provide valuable insights into this possibility.

\section*{Acknowledgments}
\noindent P.F. thanks Christos Charmousis and Mokhtar Hassaine for useful discussions and constructive feedback on the manuscript. P.F. is also grateful to the authors of Ref. \cite{buenoRegularChargedBlack2025} for useful discussions. This work is funded by the Deutsche Forschungsgemeinschaft (DFG, German Research Foundation) under Germany’s Excellence Strategy EXC 2181/1 - 390900948 (the Heidelberg STRUCTURES Excellence Cluster).

\bibliography{biblio.bib}

\end{document}